\documentclass[sort&compress,final]{aipproc}
\layoutstyle{6x9}
\begin{document}
\title[Mode Conversion around a Null Point]{MHD Mode Conversion \\around a 2D Magnetic Null Point}
\author{A.~M.~D.~McDougall}{
  address={School of Mathematics and Statistics, University of St Andrews, St Andrews, KY16 9SS, UK},
}
\author{A.~W.~Hood}{
  address={School of Mathematics and Statistics, University of St Andrews, St Andrews, KY16 9SS, UK},
}
\begin{abstract}
Mode conversion occurs when a wave passes through a region where the sound and Alfv\'en speeds are equal.  At this point there is a resonance, which allows some of the incident wave to be converted into a different mode.  We study this phenomenon in the vicinity of a two-dimensional, coronal null point.  As a wave approaches the null it passes from low- to high-$\beta$ plasma, allowing conversion to take place.  We simulate this numerically by sending in a slow magnetoacoustic wave from the upper boundary; as this passes through the conversion layer a fast wave can clearly be seen propagating ahead.  Numerical simulations combined with an analytical WKB investigation allow us to determine and track both the incident and converted waves throughout the domain.  
\end{abstract}
\maketitle
\section{Introduction}
At the conversion region the plasma $\beta$, the ratio of the gas pressure to the magnetic pressure, is approximately equal to unity.  This layer generally lies low in the solar atmosphere, in the chromosphere.  However, if a coronal null point is considered, where the magnetic field $B=0$, there will be an area surrounding the null where $\beta\approx1$.  Thus, it is reasonable to expect mode conversion to occur as a wave approaches the null, passing from low- to high-$\beta$ plasma.  

Extensive work has been done in examining mode conversion in one-dimension.  \citet{Zhugzhda1982a} looked at the conversion of fast and slow magnetoacoustic waves propagating from high- to low-$\beta$, in a gravitationally stratified, isothermal atmosphere.  Transmission and conversion coefficients were found, using an exact solution.  \citet{McDougall2007} studied conversion of slow to fast magnetoacoustic waves propagating downwards from low- to high-$\beta$.  Using a method developed by \citet{Cairns1983} transmission and conversion coefficients were found.  This method has the advantage that an exact analytical solution need not be known.  This was extended to include a non-isothermal atmosphere \citep{McDougall2008}, where it was found that the same conversion and transmission coefficients apply.  Mode conversion around a two-dimensional, magnetic null point has also been investigated \citep{McLaughlin2006}.  This looked at a fast wave propagating towards the null point from above, and focused on how the proximity of the $\beta\approx1$ layer affects the competing refraction and conversion effects.  

We look at driving a slow wave pulse along the field lines from above.  This travels from low- to high-$\beta$ towards the null point, and conversion is observed as it crosses the $c_s=v_A$ layer.  We use the WKB method to track the incoming slow wave, and as this hits the conversion region we also track the fast wave which propagates ahead.  
\section{Model}
\begin{figure}
\includegraphics[scale=0.4]{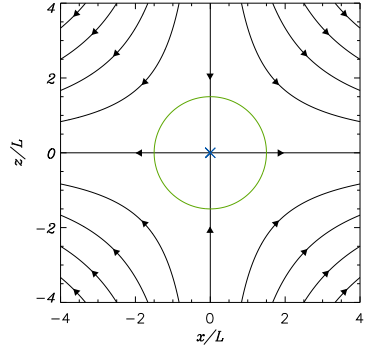}
\caption{Equilibrium magnetic field with a magnetic null point situated at the origin, denoted by a cross.  The circle shows where the sound and Alfv\'en speeds are equal.}
\label{model.fig}
\end{figure}
We use the ideal MHD equations with gravity neglected.  The equilibrium magnetic field is given by ${\bf B}_0=B_0\left(x,0,-z\right)/L$, and is shown in Figure~\ref{model.fig}.  The squared sound and Alfv\'en speeds are given by $c_s^2=\gamma p_0/\rho_0$ and $v_A^2=B_0^2\left(x^2+z^2\right)/\left(\mu\rho_0L^2\right)$ respectively, where $\gamma$ is the ratio of specific heats, $p_0$ and $\rho_0$ are the constant equilibrium pressure and density, $\mu$ is the magnetic permeability, and $L$ is a typical coronal length scale.  

The ideal MHD equations may be linearised and combined to give a pair of linear wave equations:
\begin{equation}
\frac{\partial^2v_x}{\partial t^2}=c_s^2\left(\frac{\partial^2v_x}{\partial x^2}+\frac{\partial^2v_z}{\partial x\partial z}\right)+\frac{B_0^2z}{\mu\rho_0L^2}\left[z\left(\frac{\partial^2v_x}{\partial x^2}+\frac{\partial^2v_x}{\partial z^2}\right)+x\left(\frac{\partial^2v_z}{\partial x^2}+\frac{\partial^2v_z}{\partial z^2}\right)\right],
\label{wavx.eqn}
\end{equation}
\begin{equation}
\frac{\partial^2v_z}{\partial t^2}=c_s^2\left(\frac{\partial^2v_x}{\partial x\partial z}+\frac{\partial^2v_z}{\partial z^2}\right)+\frac{B_0^2x}{\mu\rho_0L^2}\left[x\left(\frac{\partial^2v_z}{\partial x^2}+\frac{\partial^2v_z}{\partial z^2}\right)+z\left(\frac{\partial^2v_x}{\partial x^2}+\frac{\partial^2v_x}{\partial z^2}\right)\right].
\label{wavz.eqn}
\end{equation}
We define the velocity parallel and perpendicular to the magnetic field as $v_{\parallel}=\left(xv_x-zv_z\right)/\sqrt{x^2+z^2}$ and $v_{\perp}=\left(zv_x+xv_z\right)/\sqrt{x^2+z^2}$ respectively.  We may then drive a slow wave pulse exactly along the field lines by setting $v_{\perp}=0$ and $v_{\parallel}=\sin{\omega t}$, where $\omega$ is the driving frequency.  As the pulse is curved compared to our horizontal boundary, we calculate when each point hits the boundary, so the pulse is released along the boundary at the correct time.  
\section{WKB Method and Simulations}
Starting with the wave equations~(\ref{wavx.eqn}) and~(\ref{wavz.eqn}), we write $v_x=a\exp{\left(i\phi\left(x,\,z,\,t\right)\right)}$ and \break$v_z=b\exp{\left(i\phi\left(x,\,z,\,t\right)\right)}$ under the assumption that $\phi\gg1$.  We then obtain the quadratic
\[\mathcal{F}=\frac{1}{2}\left[\omega^4-\left(c_s^2+v_A^2\right)\left(p^2+q^2\right)\omega^2+c_s^2\left(p^2+q^2\right)\left(xp-zq\right)^2\right]=0,\]
\begin{figure}
\includegraphics[scale=0.9]{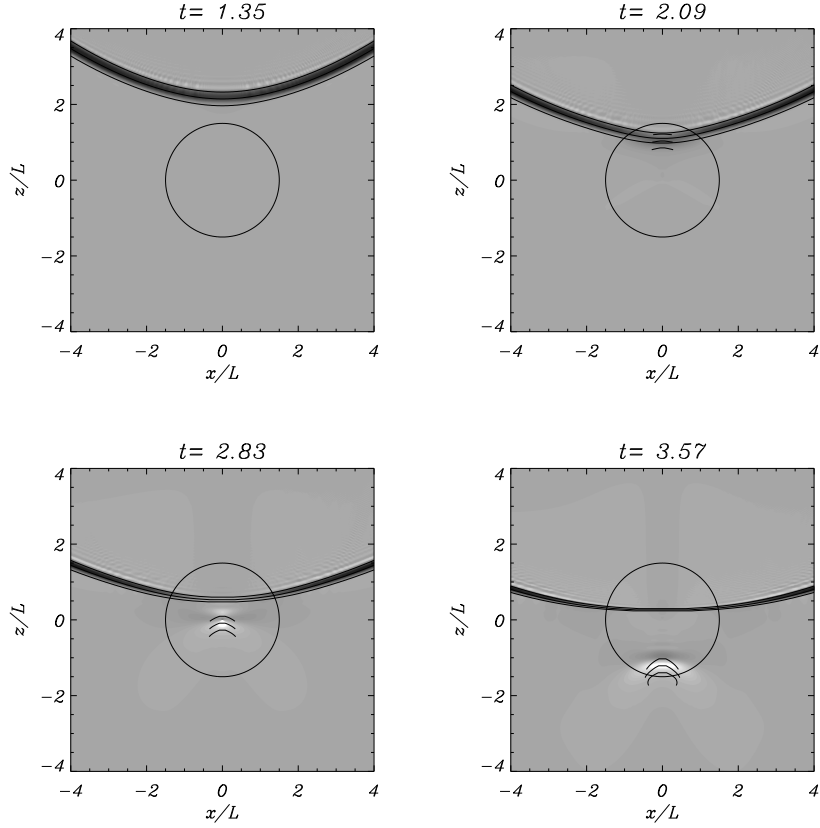}
\caption{Contour plots of the parallel velocity from the numerical simulation.  The circle denotes where $c_s=v_A$, and the cross is the location of the null point.  The lines show the front, middle and back of the slow wave and fast wave as predicted by the WKB method.}
\label{sims.fig}
\end{figure}
which we have set equal to $\mathcal{F}$, where $p=\partial\phi/\partial x$ and $q=\partial\phi/\partial z$.  The roots of this quadratic are
\[2\omega^2=\left(c_s^2+v_A^2\right)\left(p^2+q^2\right)\pm\sqrt{\left(c_s^2+v_A^2\right)^2\left(p^2+q^2\right)^2-4c_s^2\left(p^2+q^2\right)\left(xp-zq\right)^2},\]
where the plus sign gives the fast wave solution, and the minus sign the slow wave.  

Using Charpit's relations we may then obtain a system of ODEs governing the fast and slow magnetoacoustic waves:
\begin{eqnarray*}
\frac{{\mathrm{d}}p}{{\mathrm{d}}t}&=&\frac{x\left(p^2+q^2\right)}{2\omega}\pm\frac{\left[x\left(p^2+q^2\right)^2\left(c_s^2+v_A^2\right)-2pc_s^2\left(p^2+q^2\right)\left(xp-zq\right)\right]}{2\omega\sqrt{\left(c_s^2+v_A^2\right)^2\left(p^2+q^2\right)^2-4c_s^2\left(p^2+q^2\right)\left(xp-zq\right)^2}},\\
\frac{{\mathrm{d}}q}{{\mathrm{d}}t}&=&\frac{z\left(p^2+q^2\right)}{2\omega}\pm\frac{\left[z\left(p^2+q^2\right)^2\left(c_s^2+v_A^2\right)+2qc_s^2\left(p^2+q^2\right)\left(xp-zq\right)\right]}{2\omega\sqrt{\left(c_s^2+v_A^2\right)^2\left(p^2+q^2\right)^2-4c_s^2\left(p^2+q^2\right)\left(xp-zq\right)^2}},\\
\frac{{\mathrm{d}}x}{{\mathrm{d}}t}&=&-\frac{p\left(c_s^2+v_A^2\right)}{2\omega}\mp\\&\mp&\frac{\left[p\left(p^2+q^2\right)\left(c_s^2+v_A^2\right)^2-2pc_s^2\left(xp-zq\right)^2-2xc_s^2\left(p^2+q^2\right)\left(xp-zq\right)\right]}{2\omega\sqrt{\left(c_s^2+v_A^2\right)^2\left(p^2+q^2\right)^2-4c_s^2\left(p^2+q^2\right)\left(xp-zq\right)^2}},\\
\frac{{\mathrm{d}}z}{{\mathrm{d}}t}&=&-\frac{q\left(c_s^2+v_A^2\right)}{2\omega}\mp\\&\mp&\frac{\left[q\left(p^2+q^2\right)\left(c_s^2+v_A^2\right)^2-2qc_s^2\left(xp-zq\right)^2+2zc_s^2\left(p^2+q^2\right)\left(xp-zq\right)\right]}{2\omega\sqrt{\left(c_s^2+v_A^2\right)^2\left(p^2+q^2\right)^2-4c_s^2\left(p^2+q^2\right)\left(xp-zq\right)^2}}.
\end{eqnarray*}
We solve these using a fourth order Runge-Kutta scheme, allowing the position of the wave pulses to be tracked throughout the domain, as shown in Figure~\ref{sims.fig}.

The slow wave is tracked as it enters the domain and approaches the conversion layer.  Once the front of the pulse reaches this point, we begin to follow the fast wave pulse that is created.  This is only tracked for points which pass through the conversion layer, hence a much smaller portion of the wave front is followed.  
\section{Conclusions}
We have studied a two-dimensional mode conversion layer situated around a magnetic null point.  A slow wave pulse is driven along the field lines on the upper boundary, and as it hits the conversion region, where $c_s=v_A$, some of its energy is transferred to the fast mode, which we see propagating out ahead of the slow mode pulse.  The WKB method is then used to predict the position of both wave fronts as they travel through the domain.  These predicted positions are in excellent agreement with the simulations, showing the slow wave stretching out and slowing as it approaches the null, while the fast wave propagates out in front.  

In future we plan to extend the \citet{Cairns1983} method to two dimensions, allowing the quantity of conversion and transmission to be calculated, as has been done for an isothermal \citep{McDougall2007}, and non-isothermal \citep{McDougall2008}, one-dimensional model.  
\bibliographystyle{aipproc}
\bibliography{ref}
\end{document}